\documentclass{IOS-Book-Article}

\usepackage{mathptmx}
\usepackage{color}
\usepackage{tabularx}
\usepackage{amsmath}
\usepackage{listings}
\usepackage{titlesec}
\usepackage{hyperref}
\usepackage{rotating}
\usepackage{xcolor}
\usepackage{amssymb}
\usepackage{aas_macros}
\def\hb{\hbox to 10.7 cm{}}

\begin{document}

\pagestyle{headings}
\def\thepage{}
\newcommand{\linka}[1]{\href{#1}{\url{#1}}}
\begin{frontmatter}              

\title{Gadget3 on GPUs with OpenACC}

\markboth{}{October 2019\hb}

\author[A,C,B,G]{\fnms{Antonio} \snm{Ragagnin}}, %
\author[B,D]{\fnms{Klaus} \snm{Dolag}}, %
\author[E]{\fnms{Mathias} \snm{Wagner}}, %
\author[F]{\fnms{Claudio} \snm{Gheller}}, %
\author[F]{\fnms{Conradin} \snm{Roffler}}, %
\author[G]{\fnms{David} \snm{Goz}}, %
\author[C]{\fnms{David} \snm{Hubber}}, %
\author[C]{\fnms{Alexander} \snm{Arth}}, %

\address[A]{Leibniz-Rechenzentrum, M\"unchen, Germany \texttt{(antonio.ragagnin@inaf.it)}}
\address[C]{Excellence Cluster Universe, M\"unchen, Germany}
\address[B]{University Observatory of Munich, M\"unchen, Germany}
\address[G]{INAF - OATs, Trieste, Italy}
\address[D]{Max-Planck-Institut f\"ur Astrophysik, Garching, Germany}
\address[E]{NVIDIA GmbH, W\"urselen, Germany}
\address[F]{CSCS-ETH, Lugano, Switzerland }

\begin{abstract}
We present preliminary results of a GPU porting of all main Gadget3 modules (gravity computation, SPH density computation, SPH hydrodynamic force, and thermal conduction) using OpenACC directives.
Here we assign one GPU to each MPI rank and exploit both the host and accellerator capabilities by overlapping computations on the CPUs and GPUs: while GPUs asynchronously compute interactions between particles within their MPI ranks, CPUs perform tree-walks and MPI  communications of neighbouring particles.
We profile  various portions of the code to understand the origin of our speedup, where we find that a peak speedup is not achieved because of time-steps with few active particles.
We run a hydrodynamic cosmological simulation from the Magneticum project, with $2\cdot10^{7}$ particles, where we find a final total speedup of $\approx 2.$
We also present the results of an encouraging  scaling test of a preliminary gravity-only OpenACC porting, run in the context of the EuroHack17 event, where the prototype of the porting proved to keep a constant speedup up to $1024$ GPUs.
\end{abstract}

\end{frontmatter}

%
%
%
%
\section{Introduction}

The parallel N-body code Gadget3~\cite{2001NewA....6...79S,springel2005} is nowadays one of the most used high-performing codes for large cosmological hydrodynamic simulations~\cite{gadget2014}.
Gadget3 exploits hybrid MPI$/$OpenMP parallelism.
Each MPI task owns a region of the domain composed by contiguous  chunks of Hilbert-ordered particles, and, at each time-step communicates guest particles that interact with regions belonging to other MPI tasks.
Dark matter, gas and stars are sampled by particles and interact through gravity using the Barnes-Hut~\cite{barnes324hierarchical} approximation for short-range interactions and Particle-Mesh for long range interactions.
Hydrodynamics of gas particles is modelled using  an improved version of Smoothed Particle Hydrodynamics (SPH)~\cite{gingold1977smoothed} by~\cite{2016MNRAS.455.2110B}.

SPH is implemented in Gadget3 with two different modules: the first one  computes particle densities by multiple iterations and the second one computes  hydrodynamic forces.
Additionally Gadget3 implements other  physical processes as thermal conduction and sub-resolution models for star formation and black hole evolution.
Star particles interact only through gravity, however, they are created based on   properties of gas particles as in ~\cite{2003MNRAS.342.1025T}.

Each  of the above mentioned modules share the same pattern: they compute the force acting on a list of active particles by performing a tree walk over groups of one or more active particles and find all neighbouring particles within a given distance~\cite{2016pcre.conf..411R}.
Being Gadget3 a parallel code, when searching for neighbours, the code will also identify regions of the tree that belongs to a different MPI rank. 
After this identification, MPI ranks will exchange neighbouring particles.
In the second phase of each module, MPI ranks will compute forces acting on guest particles due to local contributions.
When a MPI rank has computed the interactions over the received guest particles, it will send the results back to their original MPI rank, which will merge the received contributions with the one from local neighbours.
Gadget3 uses a relatively small exchange buffer ($\approx 300MB$ per node) compared to the memory occupied by particles in a large simulation ($> 20GB$ per node), and for this reason,
it is not possible to exchange boundary particles all at one time:  first and second phases must be repeated until all active particles have been processed.

\begin{figure}
   \centering
    \includegraphics[width=\textwidth]{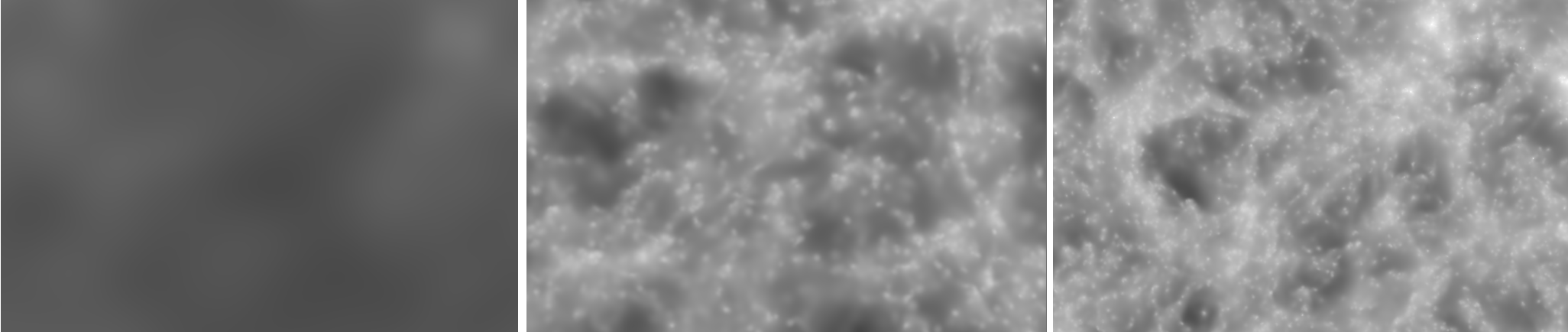}
\label{fig:g}
 \caption{Gas projected density of a portion of the cosmological simulation Magneticum Box4/hr ($2\cdot10^7$ particles). Left panel shows the gas distribution of the initial conditions of the simulation; central panel shows particles in the middle of a simulation (where dark matter haloes start forming); right panel show particles at the end of the simulation. Color values are log-scaled.}
\end{figure}

Figure \ref{fig:g} shows the projected gas density in three different phases of a simulation. 
While the initial conditions of a simulation (left panel) contains nearly homogeneous matter distribution, as simulation time increases (from left to right panels), dark matter forms haloes and filaments.
Particles inside a clustered region are driven by much stronger accelerations than particles outside these regions, thus Gadget3 uses an adaptive time-stepping scheme where active particles are updated with a kick-drift-kick solver\cite{springel2005}.

Since small time-steps have only few active particles, to improve the performance during the neighbour search, the code    drifts only tree-nodes and non-active particles that are encountered during such neighbour search.
The previously mentioned drift and filling of the export buffer are not thread-safe and are encapsulated inside OpenMP critical regions.

The most time consuming modules are Gravity ($\approx 15\%,$  of the time), SPH ($\approx 30\%$  of the time) and thermal conduction ( $\approx 14\%.$  of the time).
The remaining time is mostly taken by the domain decomposition ($\approx 16\%$ of the time) and the so called halo finder ~\cite{2009MNRAS.399..497D} ($\approx 8\%$ of the time).

In this work we present a porting of all main Gadget3 modules (gravity computation, SPH density computation, hydrodynamic force, and thermal conduction)  on GPUs using OpenACC~\cite{farber2016parallel}. 

We decided to port these modules, because, besides being the most time consuming modules, they are the ones that spend most time in loops of kernel functions, and thus are suitable for GPUS.

Our approach overlaps computations between the host and the CPU. 
GPUs asynchronously compute physical interactions between particles within the same computing node while CPUs perform tree walks, fills the export buffer and communicates particles.
Because of memory limitations,  we offload   a module per time to the GPU. 
We test our porting using the Magneticum\footnote{\linka{http://www.magneticum.org}} suite of simulations. In particular we use Box4/hr with $2\cdot10^{7}$ particles and Box3/hr with $3.8\cdot10^{8}$ particles.
We also test our code with different architectures, as P100 and V100 GPUs with NVLink technology~\cite{foley2017ultra}, with the PGI compiler with and without CUDA~\cite{nvidia2011nvidia} Unified Memory~\cite{negrut2014unified}.

In Section \ref{sec:challenges} we discuss the obstacles that prevent an easy porting of the Gadget3 code and our choices of GPU porting.
In Section \ref{sec:profilin} we profile the code and show the speedup of our porting over different portions of a cosmological hydrodynamic  simulation.
In Section \ref{sec:conclu} we draw our conclusions and discuss future projects. 

%
%
%
%
\section{Challenges and Strategies in Accelerating Gadget3}
\label{sec:challenges}
Here below we list  various limitations that  prevent an easy porting  of the whole code Gadget3 to the GPUs:

\begin{itemize}
 \item The code do not benefit from vectorisation because it stores data in  arrays of large data structures ($\approx500B$ each) that do not fit modern architecture caches. Changing the data layout to a structure of arrays would require a massive refactoring effort and introduce additional memory movement (of packing and unpacking data) in the  domain decomposition.
\item The use of  blocking MPI communications (to exchange neighbouring particles between MPI ranks)  poses a limit in fully utilising GPUs and CPUs.
\item  Time-steps with too few active particles won't fully exploit GPU parallelism, thus preventing the code to speedup;
  \item There are  thread-locking operations at each tree
  walk (drift of particles and  fill of  shared export buffer for communications).
  \item GPUs memories have less capacity than their host memories, thus simulations that keeps all data in GPUs  will require  more computing nodes than  CPU only runs.
\item Gadget3 has been built over a decennial effort of developers who implemented various flavours of gravity, SPH solvers, and sub-resolution models  that have been extensively tested; rewriting these modules using CUDA/OpenCL languages  would imply a massive rewrite of portions of such modules with  associated risks of adding mistakes.
\end{itemize}

For these reasons, a directive-based approach that uses OpenACC \cite{farber2016parallel} has been adopted. 
This reduces modifications of the ongoing development of Gadget3 and furthermore makes it possible to still run the code on CPU-only systems.



\subsection{Memory Transfer}

To minimize communication between CPUs and GPUs, one  would ideally load the initial conditions of the simulation in the memory of the GPU and  run the whole simulation on  GPUs.
This solution has two problems: first of all,   time steps with few active particles won't perform on the GPUs, and since current GPUs typically have less memory than their hosts, one would need more nodes than a CPU-only run.

To clarify the last point, let's  consider the case of a very large cosmological simulation that was run within the LRZ Extreme Scaling Workhop in 2015~\cite{2016arXiv160901507H}. Such simulation (Magneticum Box0/mr)  had $1.2\cdot10^7$ particles per node, each node  was allocating  $4GB$ for the Barnes Hut tree,  $22GB$ for the basic quantities used in gravity (e.g. position, mass, acceleration ecc..), and additional $14GB$ for the  SPH-only part (that is split in density computation and hydro-force computation), $0.6GB$
for the metal evolution and an additional amount of $4GB$ for the active particle list and to store the Hilbert space-filling-curve keys, for a grand total of $40GB$ per node. 

It is clear that a $16GB$ GPU system (as for instance, the ones in Piz Daint\footnote{\linka{https://www.cscs.ch/computers/piz-daint/}})  would not be able to store the same number of particles of its underlying host.
On the other hand, it has enough memory to store  the particle properties of each single Gadget3 module at a  time.

To solve this issue and to be able to exploit the GPU memory at its best, we decided to only upload, for each Gadget module,  the properties that are necessary for such module (or for other successive modules) in the current time step.
With this technique we are able to upload more particles per timestep, but we can upload only the minimal set of properties required by each module at time.
The drawback of this approach is that at each timestep we need to download the data back to the GPU, with its associated overhead.

To further minimise the data transfer of the particle properties, we send separately
properties that are read-only  (masses, positions,  ecc..) and download only updated properties  (e.g. acceleration).

Additionally, with this approach we minimise the amount of code we have to write/modify: we use the same Gadget routines used to process guest neighbouring particles coming from a different MPI rank.
We set up the code so  GPUs use the already existing routines to exchange data, but in this case, particles are exchanged between host and GPUs.

\subsection{Adaptive Timesteps}

Large, high resolution, cosmological simulations have both void regions and clustered regions.
Particles in void regions evolve with large timesteps because of the small force acting on them, compared to clustered regions where the stronger force requires very small timesteps.

After nearly half of the simulation time, it is very common to have timebins
with only one or very few active particles.
Since time-steps with such a low amount of active particles won't benefit from the single instruction multiple thread (SIMT) paradigm of GPUs, we decided to keep small timebins (with less than a given threshold $N_{min}$ active particles) to run on the CPU only.

The reason behind this choice is twofold: (i) with a high number of active particles, the offload time is small compared to computations and (ii) it is possible to drift all particles and tree-nodes of the simulated volume at the beginning of these time-steps in the host, with OpenMP.

OpenACC turned out to be the best tool to implement this decision because it makes it possible to use the same code on both GPU and CPU with a small effort.

\subsection{MPI Communication}

One of the main advantages  of our porting is that it overlaps CPU work with the GPU computation.
We decided to overlap the CPU and the GPU computation in the following way:
while the GPU loops over the active particles and computes local interactions,
the CPU takes care of walking the tree for each active particle in order to perform all MPI send/receive of guest particles.

When the host receives a list of guest particles it decides to queue it to the GPU computation or to process it on the CPU, based on the facts that (i) the GPU did finish local interactions or not and (ii) the number of received particles is less than  $N_{min}$.

\subsection{Barnes-Hut, SPH and Thermal Conduction Differences}

Although SPH, thermal conduction and Barnes Hut algorithms have many similarities,  there are some main differences to take into account when porting Gadget on GPUs.
First of all, in a Barnes-Hut solver,  particles interact with distant tree nodes as they were point-like pseudo particles, in contrast with SPH and conduction solvers where there are only  particle-particle interactions within a pre-defined distance.
As a consequence, the implementation  of Barnes-Hut algorithm embed the particle-particle interaction computation in the tree walk itself.
On the other hand, in SPH and thermal conduction solvers, neighbours are collected in a list and processed in a separate step.
Additionally, SPH and thermal conduction need to find a set of neighbours within a fixed distance, while Barnes-Hut operates with a so-called  opening criteria, namely the angle between the target particle and the tree cells.

In our OpenACC porting, this implies that gravity acceleration computation will be inside a tree walk branch, which will limit the peak GPU performance. 
While in the SPH and conduction modules it is possible to disentangle the tree walk from the force computation.
The drawback is that it is not well known a priori the amount of neighbours of
a given  SPH particle (especially in zoom-in simulations).
The CPU implementation overcomes this problem by allocating a neighbour buffer
for each thread of a size that is equal to the number of local particles. 
Since it is practically impossible to allocate such a long buffer on each GPU thread, our porting performs a tree walk and neighbour interactions in chunks of $N_{chunk}$ neighbours.

%
%
%
%
\section{Profiling}
\label{sec:profilin}

We  tested  our implementation over different setups and architectures, were we  found the values of  $N_{min}=10^3$ and $N_{chunk}=32$  to be optimal in always maximizing the speedup.
Time steps with a number of active particles less than $10^3$ tipically performs better in the CPU than in the GPU.

The value of $N_{min}$ do not need to be extremely accurate.
In fact the number of particles between one time bin and the other changes exponentially, from our experience, the number of particles between a small step and the next one goes from few hundreds to few thousands.


\subsection{Tests of One Density Iteration}
\label{ssec:oneiter}

\begin{figure}
   \centering
    \includegraphics[width=\textwidth]{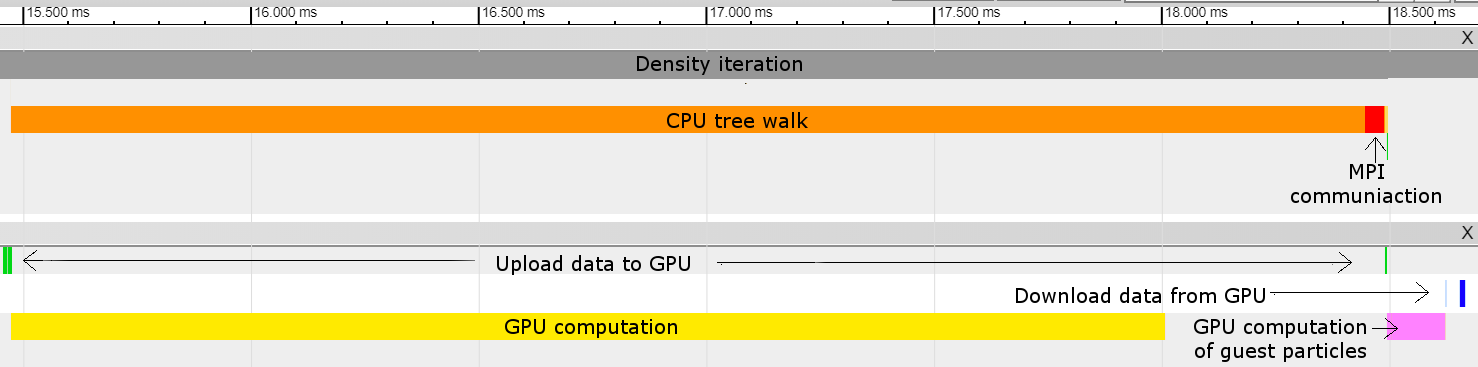}
\label{fig:profilong}
 \caption{Timeline of the profiling obtained with the nvToolsExt library of the OpenACC code: upload and download to GPU (green and blue bars), GPU computation (yellow and purple bars), CPU tree walk (orange bar) and MPI communications (red bar). The full iteration takes $2.4s,$ while the CPU version of the code, run with the same setup, took $11.4s.$}
\end{figure}

Figure \ref{fig:profilong} shows a time-line of the profiling (obtained with the nvToolsExt library\footnote{\linka{https://docs.nvidia.com/gameworks/content/developertools/desktop/nsight/nvtx_library.htm}})  of a SPH density  iteration over all particles of  the Magneticum/Box4/hr simulation ($2\cdot10^{7}$ particles) on a Power9 system with $2$ MPI ranks per node, each with $20$ OpenMP threads plus one Tesla V100 GPU.
With this setup we used all cores of a node (and without using hyper-threading).
Each socket has NVLink interconnection technology between CPUs and GPUs.

In this setup, upload and download timings (green and blue bars) sums up to $0.053s$ (for a total of $1.2GB$) and are negligible compared to the GPU computation time (yellow and purple bars), that take up to $2.4s$.
CPU   tree walks (orange bar) takes $2.9s$  and MPI communications (red bar) take $0.04s$ and overlaps the GPU computations.

The whole density iteration took $3.2s,$ while the same set-up, when run completely on CPUs, took $11.4s.$
Of which $11s$ spent in computation and the remaining $0.4s$ spent in MPI communications.

Thus, a SPH density iteration over all active particle have a speedup of $4.5.$
However, the speedup of the full simulation will be lower because a number of iterations have only very few active particles and do not perform well on GPUs.

We then briefly tested the possibility of using Unified Memory  for our OpenACC porting.
In particular, we run Magneticum Box4/hr simulations with 2 MPI rank, each using one V100 GPU connected with NVLink.

The iteration without Unified Memory took  $1.9s$, where  $0.4s$  were  spent  in  memory transfer, while the run with Unified Memory: $2.0s.$
They pratically takes the same time.
The advantage of Unified Memory is that one does not have to manually write code to restrict the transfer of data to its minimum necessary amount.

\subsection{Tests of One Timestep}
\label{ssec:timestep}
 
\begin{figure}
   \centering
    \includegraphics[width=0.83\textwidth]{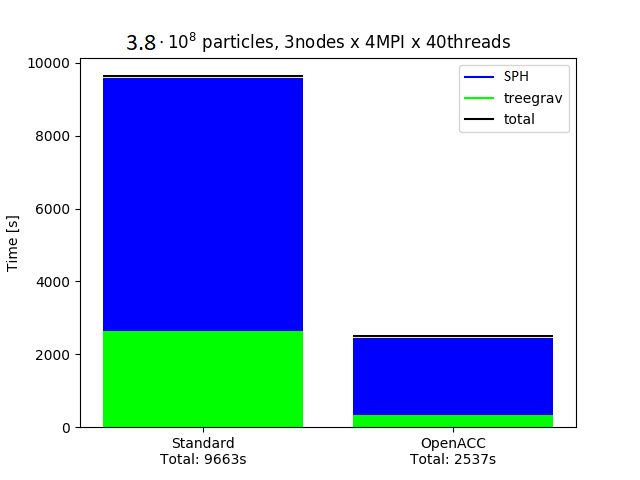}
\label{fig:modules}
 \caption{Time consumed by  SPH and gravity for a simulation with  $3.8\cdot10^8$ particles. Left panel shows the time consumed by the standard version, run over 3 nodes, each with 4 MPI ranks and each with 40 threads. Right panel shows the same configuration for the OpenACC porting where each MPI rank had one Tesla V100 GPU connected with NVLink. }
\end{figure}

Figure \ref{fig:modules} shows the timing of a whole timestep of the various Gadget modules.
In this test, to maximize the number of particles for a given node, we run  Magneticum Box4/hr simulation, that has $3.8\cdot573^3=1.2\cdot10^8$ particles.
We run this simulation on a Power9 system with $4$ sockets per node, each with $10$ cores and one V100 GPU. 
We used $3$ nodes, each with 4 MPI ranks, and each MPI rank with $40$ OpenMP threads (thus using Power architectures hyper threading).

Here we can see how, at least for the first time-step, most of the speedup is consistent over both the Barnes-Hut solver and the full SPH computation to a factor of $\gtrapprox3$ for SPH and $\approx 4$ for the gravitaty computations.
 
In particular, all SPH density iterations within the timestep took $1600s$ for the GPU version and $5400s$ (with a speedup of $3.3$) for the CPU version. While the SPH computation of hydrodynamic forces tooks  $200s$ for the GPU version and  $700s$ for the CPU version (with a speedup of $3.5$).

The speedup of this test case is lower than the one obtained in the previous test case because in a whole timestep there are density iterations that have a very low number of particles.


\subsection{Tests of Full Run}
\label{ssec:fullrun}

We then run a whole simulation with Barnes-Hut, SPH and thermal conduction ported with OpenACC. As described above, we offload these modules to the GPU only when the number of active particles is greater than
the treeshold $N_{min}=10^3.$

We run such simulation on the Piz Daint system.
Here we used $8$ MPI tasks in $1$ node, $4$ OpenMP threads and one Tesla P100 for each MPI task. In such system, GPUs are connected to the host with PCI Express  technology. At the end of a simulation, the speed-up (compared to the same set-up of the Gadget3 standard version) are as follows:

\begin{itemize}
 \item Barnes-Hut speedup:  $1.8$
 \item SPH speedup:  $2.6$   
 \item Thermal conduction speedup: $3.0$
 \item Total speedup: $2.1$
\end{itemize}

Noteworthy, SPH speedup is lower than the speedup obtained for a single timestep as in the previous sub section (and for a single iteration of a density computation).
The reason behind this slowdown is twofold: the number of density iterations that contains a small number of active particles increases as the simulation time evolves. For this reason we found a final total speedup to be lower than the one obtained for the first timestep.

After porting Gravity, SPH and thermal conduction to the GPU, one of the upcoming bottlenecks became the cooling and star formation module, taking $\approx5\%$ of the computing time.
Here we upload the cooling tables to the GPUs and keep them there as long as needed. We then run the whole cooling and star formation process in the GPU, which have a speedup of $\approx 1.6,$ when comparing a run with P100 GPUs and a run with 12 Haswell CPUs\footnote{\linka{https://www.cscs.ch/publications/stories/2018/conradin-roffler-my-internship-at-cscs/}}.

\subsection{Scaling Test of Gravity Only}
\label{ssec:scaling}

\begin{figure}
   \centering
    \includegraphics[width=0.85\textwidth]{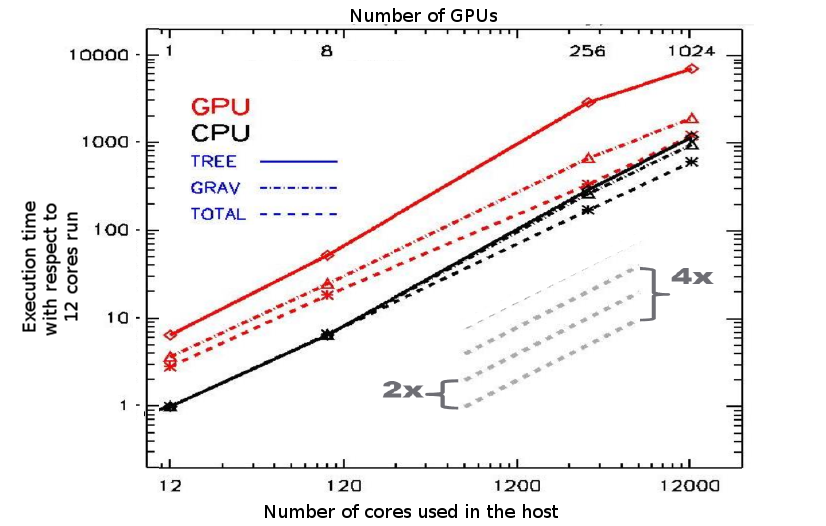}
 \caption{Scaling of a preliminary Gadget3 OpenACC porting of the gravity module, done at the EuroHack17 at CSCS. Y-axis show the speedup with respect to the CPU only version over 12 cores. Black (bottom) lines show the data by varying the number of MPI ranks (X axis) for a CPU-only run. Red (upper) lines show data for the OpenACC version, on the same number of CPUs and one additional GPU for each MPI rank. Continuous lines show the tree-walk speedup, dotted-dashed lines show the speedup for the gravity module, dashed lines show the speedup of the whole time step.}
\label{fig:gpu_scaling}
\end{figure}

Figure \ref{fig:gpu_scaling} shows the results of a scaling obtained at the EuroHack17 at CSCS\footnote{\linka{https://github.com/fomics/EuroHack17/wiki/GadgetACC}}.
At that time the preliminary version of the code was able to run over one GPU per computing node, and we ported only the  first phase of the Barnes-Hut gravity solver.
In this test we run a gravity-only run with increasing particle sizes in order to occupy more and more computing nodes, and varied the number of $MPI ranks$ up to $1024.$
Where the data point with the largest number of CPUs (and GPUs) is simulation is  Magneticum Box2/hr, that has with $2\cdot1584^3=7.9\cdot10^9$ particles.
Both the OpenACC and the standard runs uses the same amount of CPUs.

%
%
%
%

\section{Conclusions and Outlook}
\label{sec:conclu}
We presented a porting of all main Gadget3 modules (gravity computation, SPH density computation, hydrodynamic force, and thermal conduction)  on GPUs using OpenACC.

We justified our choices of the porting as:
\begin{itemize}
    \item  the use OpenACC  minimizes the rewriting of code and to let the community keep working on both CPU and GPU;
    \item  OpenACC is also useful since we offload to the GPU only timesteps with a high number of active particles (as they won't perform well in a GPU);
    \item during a simulation, and at every timestep, we offload to the GPU only one module per time as to maximize the number of particle per each host;
    \item by doing so, we exploit the host machine by overlapping GPU and CPU computations (as the CPU takes care of neighbour exchanges).
\end{itemize}

We used the same kind of porting paradigm on all Gadget modules, and showed how it keeps its speedup over different architectures (e.g. V100+NVLink or P100+PCI Express) and number of devices.
This points to the direction that this kind of porting, which involves CPU/GPU computational overlap, is stable over different modules and architectures and may be useful for other multi-node N-body solvers.

Although we performed only one test that executes all modules up to the end of the simulation (Sec. \ref{ssec:fullrun}), the various tests gave us the possibility to probe the performance on different configurations: with V100+NVLink technology (Sec. \ref{ssec:oneiter}), with P100+PCI Express (Sec. \ref{ssec:timestep}), and over a large number of GPUs (Sec. \ref{ssec:scaling}).
The EuroHack17 scaling in particular,  showed how our approach (although it tests only one module, namely the gravity module) is capable of keeping its speedup up to a thousand of GPUs.

These tests were also useful to investigate the origin of the speedup by gradually increasing the profiled region of simulations, here we found that: (i)  a single SPH density iteration, where we found a speedup of $\approx4.5$; (ii)  a full time step, where we found a timestep of $\approx 3.5;$ (iii) to a  full simulation and to a large number of GPUs where we found a total speedup of $\approx 2.$

We briefly tested Unified Memory and found that, in our preliminary tests, this technology 
reaches the same performance of our explicit memory management.
Unified Memory is a solution we will explore further because one does not have to manually set up the data transfer (which is not trivial in Gadget, since every timestep has only a subset of active particles).

Additionally, from that experience we found that the Domain Decomposition and the Tree Build are the new bottleneck of very large runs, once one speeds up the other modules with our OpenACC porting.

An initial step towards porting other modules of Gadget have been done, where we ported the cooling and star formation module. The other upcoming bottlenecks are the domain decomposition and the tree build algorithms, which by now are neither MPI parallel nor OpenMP parallel.

\section*{Acknowledgement}
This work was carried out within the EuroEXA and ExaNeSt (FET-HPC) project (grant no. 754337 and no. 671553).  We thank Emmanouil (Mano) Farsarakis from EPCC; Margarita Petkova and Rupam Bhattacharya from TUM, Milena Valentini from LMU;  Luigi Iapichino, Nicolay Hammer, and Michele Martone from LRZ.

%
%
%
%

\bibliographystyle{ieeetr} 
\bibliography{ref}

\end{document}